\documentclass[a4paper]{jpconf}
\usepackage{amsmath,amssymb,epsfig,bm,mathrsfs,color}
\usepackage{slashed}
\newcommand{\be}{\begin{equation}}
\newcommand{\ee}{\end{equation}}
\newcommand{\bea}{\begin{eqnarray}}
\newcommand{\eea}{\end{eqnarray}}
\newcommand{\half}{\frac1 2}

\newcommand{\apj}{Astrophys. J.}
\newcommand{\nat}{Nature}
\newcommand{\prc}{Phys. Rev. C}

\newcommand{\de}{\partial}

\newcommand{\vectau}{{\bm \tau}}
\newcommand{\vecrho}{{\bm \rho}}

\definecolor{red}{rgb}{0.8,0,0}
\definecolor{orange}{rgb}{0.8,0.2,0.0}
\definecolor{blue}{rgb}{0.3,0.0,0.8}

\begin{document}

\title{Equation of state of hypernuclear matter: tuning 
 hyperon--scalar-meson couplings}

\author{Giuseppe Colucci and Armen Sedrakian}
\address{Institute for Theoretical Physics, J.~W.~Goethe-University, D-60438 Frankfurt-Main, Germany}
\ead{colucci@th.physik.uni-frankfurt.de,
sedrakiani@th.physik.uni-frankfurt.de}

\begin{abstract}
  We discuss to which extent the modifications of the
  hyperon--scalar-meson coupling constants affect the equation of
  state (EoS) hypernuclear matter. The study is carried out within a
  relativistic density functional theory. The nucleonic matter is
  described in terms of a density-dependent parametrization of
  nucleon-meson couplings, whereas the hyperon--meson couplings are
  deduced from the octet model.  We identify the parameter space of
  hyperon-meson couplings for which massive stellar configurations
  with $M\le 2.25 M_{\odot}$ exist. We also discuss the EoS at finite
  temperatures with and without of a trapped neutrino component and
  show that neutrinos stiffen the EoS and change qualitatively the
  composition of stellar matter.
\end{abstract}

\section{Introduction}\label{sec:1}

The recent observations of two-solar-mass pulsars in binary orbits
with white dwarfs~\cite{2010Natur.467.1081D,2013Sci...340..448A}
spurred an intensive discussion of the phase structure of dense
matter, which is consistent with the implied observational lower bound
on the maximum mass of any sequence of compact stars based on the
unique equation of state (hereafter EoS) of dense matter. In this
article we review and summarize the key result of our study of
hypernuclear matter in the context of these observations of massive
compact stars~\cite{CS2013}.

Large central densities achieved in massive compact stars may require
substantial population of heavy baryons (hyperons), because these
become energetically favorable once the Fermi energy of neutrons
becomes of the order of their rest mass. The onset of hyperons (and
more generally any new constituent) reduces the degeneracy pressure of
a cold thermodynamic ensemble. Therefore the EoS becomes softer than
in the absence of the hyperons (or any other constituent).  This
decreases the maximum mass of a compact stars to values which
contradict the observation of massive compact stars in nature. The
controversy between the theory and observations is the 
``hyperonization puzzle'' in compact stars.

Hyperons in dense nuclear matter have been studied using a number of
methods, including Lagrangian based relativistic density functional
methods~\cite{1982PhLB..114..392G,
  1989NuPhA.493..549W,1990PhRvL..64...13K,1991NuPhB.348..345E,1991PhRvL..67.2414G,1995PhLB..349...11E,1996PhRvC..53.1416S,weber_book,2007PrPNP..58..168S}
with coupling parameters fixed by the nuclear phenomenology, as well
as models based on hyperon-nucleon potentials (for recent work 
see~\cite{2011PhRvC..84c5801S,2012JPhCS.342a2006L}).  While the
potential models fail to produce heavy enough stars and are most
likely  accurate at densities not much larger than the nuclear
saturation density, the relativistic Lagrangian based models are
suitable candidates for extrapolation to high density regime. 

The hypernuclear EoS was investigated~\cite{CS2013} by us recently
in the framework of the density-dependent relativistic density
functional method.  In particular, we focus on the sensitivity of the
EoS of hypernuclear matter to the unknown hyperon--scalar-meson
couplings. These are constrained only by imposing SU(6) symmetry
breaking and the nonet mixing. Within this framework, we argue that
the parameters can be tuned such that two-solar massive hyperonic
compact stars can exist. The EoS and composition of matter were also
studied at finite temperatures relevant for the hot proto-neutron star
stage of evolution and it was shown that the neutrino component 
stiffens the EoS of hypernuclear matter and substantially changes the
composition of matter~\cite{CS2013}. 

\section{Theoretical model and choice of couplings}\label{sec:2}

The relativistic Lagrangian density of our model reads
\bea\label{eq:lagrangian} \nonumber {\cal L} & = &
\sum_B\bar\psi_B\bigg[\gamma^\mu \left(i\de_\mu-g_{\omega B}\omega_\mu
  - \half g_{\rho B}\vectau\cdot\vecrho_\mu\right)
- (m_B - g_{\sigma B}\sigma)\bigg]\psi_B \\
\nonumber & + & \half \de^\mu\sigma\de_\mu\sigma-\half
m_\sigma^2\sigma^2 - \frac{1}{4}\omega^{\mu\nu}\omega_{\mu\nu} + \half
m_\omega^2\omega^\mu\omega_\mu -
\frac{1}{4}\vecrho^{\mu\nu}\vecrho_{\mu\nu}
+ \half m_\rho^2\vecrho^\mu\cdot\vecrho_\mu\\
& + & \sum_{\lambda}\bar\psi_\lambda(i\gamma^\mu\de_\mu -
m_\lambda)\psi_\lambda - \frac{1}{4}F^{\mu\nu}F_{\mu\nu}, \eea where
the $B$-sum is over the $J^P = \half^+$ baryon octet, $\psi_B$ are the
baryonic Dirac fields with masses $m_B$.  The meson fields
$\sigma,\omega_\mu$ and $\vecrho_\mu$ mediate the interaction among
baryon fields, $\omega_{\mu\nu}$ and $\vecrho_{\mu\nu}$ represent the
field strength tensors of vector mesons and $m_{\sigma}$,
$m_{\omega}$, and $m_{\rho}$ are their masses. The baryon-meson
coupling constants are denoted by $g_{mB}$. The last line of
Eq.~(\ref{eq:lagrangian}) stands for the contribution of the free
leptons, where the $\lambda$-sum runs over the leptons
$e^-,\mu^-,\nu_e$ and $\nu_\mu$ with masses $m_\lambda$. The last term
is the electromagnetic energy density.  The contribution of neutrinos
is included at temperatures above those at which the neutrinos
decouple from matter, which is of order of several MeV. 

The nucleon--meson coupling constants have been taken according to the
DD-ME2 density-dependent parameterization \cite{ring2005}. The density
dependence of the couplings implicitly takes into account many-body
correlations among nucleons which are beyond the mean-field
approximation.  The nucleon-meson coupling constants are parametrized
as $g_{iN}(\rho_B) = g_{iN}(\rho_{0})h_i(x)$, for $i=\sigma,\omega$,
and $g_{\rho N}(\rho_B) = g_{\rho N}(\rho_{0})\exp[-a_\rho(x-1)]$ for
the $\vecrho_\mu$-meson, where $\rho_B$ is the baryon density,
$\rho_0$ is the saturation density, $x = \rho_B/\rho_0$ and the
explicit form of the functions $h_i(x)$ and the values of couplings
can be found elsewhere~\cite{CS2013,ring2005}. The pressure and energy
density of the model is further supplemented from the contribution
coming from the so-called rearrangement
self-energy~\cite{1995PhRvC..52.3043F}, which guarantees the
thermodynamical consistency.

In order to fix the hyperon--meson couplings we consider the
SU(3)-flavor symmetric octet model.  Due to the universal coupling of
the $\vecrho_\mu$ meson to the isospin current \cite{VMD_model} and 
the ideal mixing between the $\omega$ and $\phi$
mesons\cite{2009JHEP...07..105K}, the couplings between hyperons and
vector mesons are as follows
\be \label{vector_couplings}
\begin{split}
 g_{\Xi\rho} & = g_{N\rho}, \qquad g_{\Sigma\rho}= 2g_{N\rho}, 
 \qquad g_{\Lambda\rho} = 0,\\
 g_{\Xi\omega} & = \frac{1}{3}g_{N\omega},\quad
 g_{\Sigma\omega} = g_{\Lambda\omega} = \frac{2}{3}g_{N\omega}.
\end{split}
\ee 
Within the octet model the baryon-scalar mesons couplings of the
scalar octet can be expressed in terms of only two parameters, the
nucleon--$a_0$ meson coupling constant $g_S$ and the $F/(F+D)$ ratio
of the scalar octet~\cite{de_swart1963}. By considering the
mixing with the scalar singlet state, one is then left with the
following relation between the coupling of the baryons with the
$\sigma$-meson \cite{CS2013}: \be\label{coupling_variation}
2(g_{N\sigma} + g_{\Xi\sigma}) = 3g_{\Lambda\sigma} +
g_{\Sigma\sigma}, \ee We assume that the hyperon coupling constants
must be positive and less than the nucleon coupling constant. Then, by
solving Eq.\ (\ref{coupling_variation}) for one of the dependent
hyperon-$\sigma$ meson coupling constant, say $g_{\Xi\sigma}$,  we
obtain 
 \be \label{additional_constraints2}
g_{N\sigma}\le \half (3g_{\Lambda\sigma} + g_{\Sigma\sigma}) \le
2g_{N\sigma}.  \ee 
We further proceed by first fixing the value of
$g_{\Lambda\sigma}$ coupling constat at the value provided by the Nijmegen Soft Core
(NSC) hypernuclear potential~\cite{1989PhRvC..40.2226M}
and varying the range of couplings $g_{\Sigma\sigma}$ within the limits
provided by Eq. (\ref{additional_constraints2}) and then we repeat the
calculations by interchanging the role of $\Sigma$ and $\Lambda$
hyperons. We also studied the case where one of couplings is fixed to
the depth of the potential of the $\Sigma^-$ and $\Lambda$ hyperons in
nuclear matter at the saturation density. 

\section{Results}\label{sec:5}

\begin{figure}[t]
\centering
 \includegraphics[width=12cm]{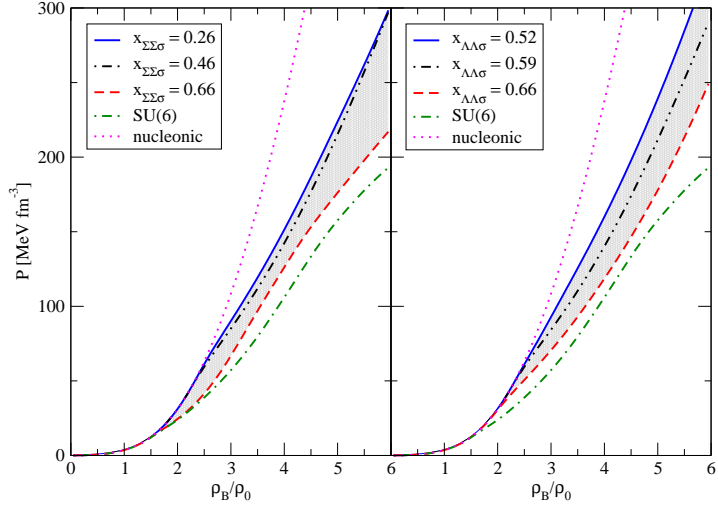}
\caption{ Equations of state of hypernuclear matter for a range of
  values of hyperon-$\sigma$ meson couplings defined in terms of
  $x_{H\sigma} = g_{H\sigma}/g_{N\sigma} $, $H\in \Lambda, \Sigma$.
  The nucleonic EoS (dotted line, magenta online) and hyperonic EoS
  with SU(6) quark model couplings (dot-dashed line, green online) are
  shown as a reference.  The nucleonic coupling constants correspond
  to the DD-ME2 parametrization~\cite{ring2005}; the hyperon-vector
  meson couplings are fixed as explained in the text.  Left panel: we
  assume $x_{\Lambda\sigma} = 0.58$, as in the NSC potential model,
  and a range $0.26\le x_{\Sigma\sigma} \le 0.66$ which generates the
  shaded area. Right panel: we assume $x_{\Sigma\sigma} = 0.448$ and a
  range $0.26 \le x_{\Lambda\sigma} \le 0.66$.  The cases
  $x_{\Sigma\sigma} = 0.46$ (left panel) and $x_{\Lambda \sigma} =
  0.59$ (right panel), shown by dash-double-dotted (black online)
  lines, fit the depth of the potentials of the $\Sigma^{-}$ and
  $\Lambda$ hyperons in nuclear matter at saturation.  }
 \label{fig:hyperons_T_0}
\end{figure}

The dependence of the EoS on the variation of the hyperon--scalar
meson coupling at $T=0$ is shown in Fig.~\ref{fig:hyperons_T_0} for a
range of parameter space (for details see the figure caption).   
It is clearly seen that the hyperonization of matter softens the
nucleonic EoS. The softening is smallest for the lowest possible
values of the couplings of the hyperons to the scalar mesons. 

\begin{figure}[tb]
 \centering
 \includegraphics[width=8.5cm]{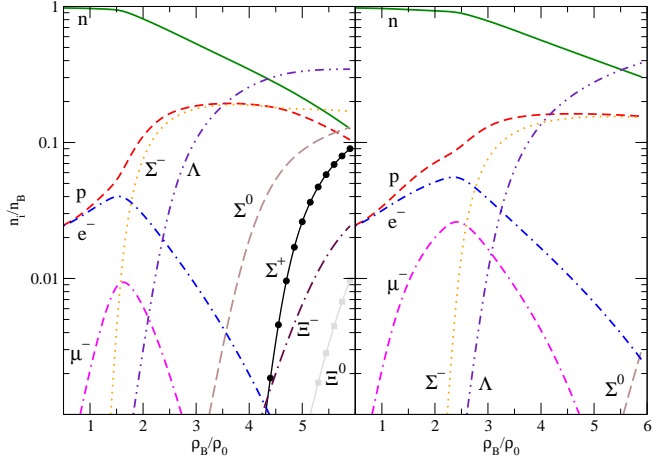}
 \caption{
Particle fractions in hypernuclear matter at $T=0 $.
 Left panel: hyperon--scalar meson couplings are fixed as in the SU(6)
 symmetric quark model. 
 Right panel: a stiff hypernuclear EoS from our parameter space with 
 $x_{\Sigma\sigma} = 0.448$ and  $x_{\Lambda\sigma} = 0.52$.}
 \label{fig:fractions_T_0}
\end{figure}

Fig.~\ref{fig:fractions_T_0} shows the particle fractions of fermions
at zero temperature, for the limiting cases of the SU(6) quark model
couplings (left panel) and the stiffest hypernuclear EoS (right
panel).  The first EoS is characterized by large hyperon--scalar-meson
couplings, whereas the second by small ones.  Thus, the larger are the
hyperon--scalar-meson couplings the more favorable is the formation of
hyperons and the softer is the EoS. The mass radius relation for these
EoS 
\begin{figure}[tb]
\begin{center}
\hspace{3.8cm}
 \includegraphics[height=9.0cm,width=12.0cm]{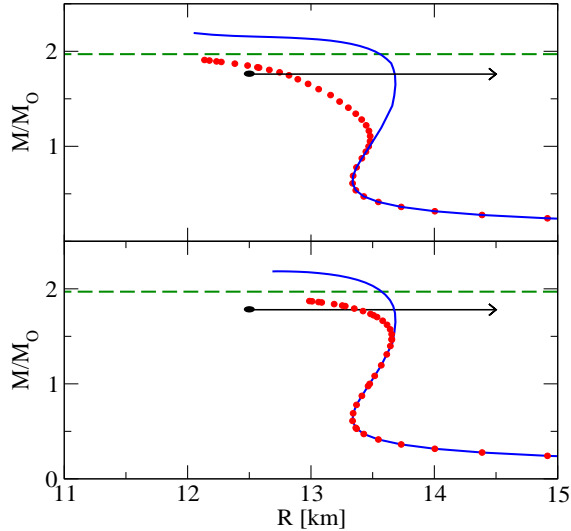}
 \caption{
The mass--radius relations for compact hypernuclear stars at zero
temperature.  
The solid (blue) lines show the cases 
$x_{\Sigma\sigma} = 0.26$ and $x_{\Lambda\sigma} = 0.58$ (upper panel)
and 
$x_{\Sigma\sigma} = 0.52$ and $x_{\Lambda\sigma} = 0.448$ (lower
panel). The (red) dots show the cases 
$x_{\Sigma\sigma} = 0.66$ and $x_{\Lambda\sigma} = 0.58$ (upper panel)
and 
$x_{\Sigma\sigma} = 0.66$ and $x_{\Lambda\sigma} = 0.448$ (lower panel).
The dash-dotted (green) line shows the observational lower 
limit on the maximum mass $1.97 M_{\odot}$. The arrow shows the 
mass-radius constraint of Ref.~\cite{2013ApJ...762...96B} at 
2$\sigma$  level, which is $M = 1.76M_{\odot}$ and $R\ge 12.5$ km.
}
 \label{fig:mass_vs_radius}
\end{center}
\end{figure}
is shown in Fig.~\ref{fig:mass_vs_radius}, which demonstrates the
large enough masses can be achieved within the present model that are
consistent with the observational lower bound on the maximum mass of a
compact star.
\begin{figure}[bt]
\centering
\vspace{0.4cm}
\includegraphics[width=12cm]{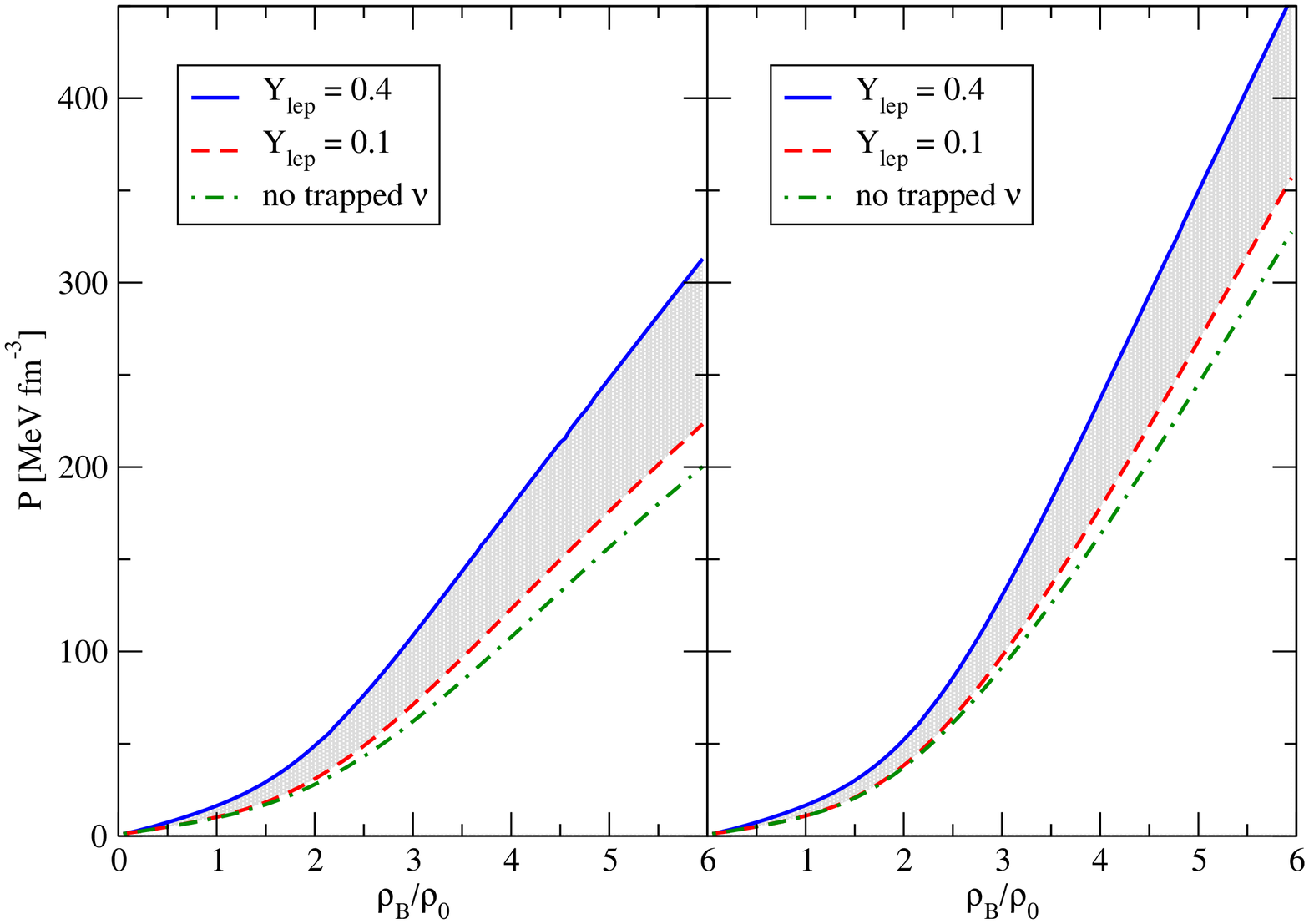}
\caption{Equation of state in the DD-ME2 parametrization
  \cite{ring2005} at finite temperature, $T=50$ MeV.  Left panel: the
  hyperon--scalar meson coupling constants are fixed by the quark
  model. In this figure the dependence on the presence of trapped
  neutrinos is shown. The dot-dashed line (green online) corresponds
  to the case without neutrinos. The presence of
  neutrinos make the EoS stiffer.  The dashed region represents the
  variation of the EoS with the lepton fraction. The dashed line (red
  online) corresponds to a lepton fraction $Y_L = 0.1$ and the
  full line (blue online) to $Y_L = 0.4$.  Right panel: same as
  left, in which the coupling constants are the one of the stiffest
  case considered, corresponding to $g_{\Sigma\sigma} =
  0.448g_{N\sigma}$, fixed by the NSC potential and
  $g_{\Lambda\sigma} = 0.52g_{N\sigma}$.}
\label{fig:hyperon_T_50}
\end{figure}
Fig.~\ref{fig:hyperon_T_50} shows the EoS of finite temperature
hypernuclear matter with trapped neutrinos; the lepton fraction is
varied in the range $0.1 \le Y_{L} \le 0.4$ (shaded area in
Fig.~\ref{fig:hyperon_T_50}).  This corresponds to the shaded
area. The left and right panels correspond to the softest and the
stiffest hypernuclear EoS described above. This compared to the
neutrino-less case ($Y_\nu\neq 0$). Clearly neutrinos stiffen the EoS
matter - the larger is the fraction of neutrinos, i.e. $Y_L$, the
stiffer is the EoS.  The stiffening of the EoS can be attributed to
the fact that the thermal population of neutrinos adds its
contribution to the pressure of matter.

The obtained EoS can be conveniently represented by piecewise polytropic 
EoS of the form~\cite{CS2013}~\footnote{We use the occasion to correct a misprint
  in the original formula (Eq.~(32) of~\cite{CS2013}), where the density
  normalization by $\rho_0$ is missing. }
\be
P = \sum_{i=1}^4 K_i (\rho/\rho_0)^{\Gamma_i}\theta(\rho-a_i\rho_0)\theta(b_i\rho_0-\rho), 
\ee
where $\Gamma_i$ is the polytropic index, $K_i$ is a dimensionful
constant, $[K_i] =$ MeV fm$^{-3}$, and $\rho_0$ is the saturation
density; the values of the fit parameters can be found  in~\cite{CS2013}.

\section{Conclusions}
\label{sec:6}

Because the information on the properties of hypernuclear matter is
far less extensive than for nucleons it is currently impossible to
exclude hyperons as constituents of densest regions of compact stars.
Our study \cite{CS2013} confirms this within a specific relativistic
density functional approach to hypernuclear matter with tuned
hyperon--scalar-meson couplings. We find that hyperonization in
massive stars is favored for small ratios of the
hypernuclear-to-nuclear couplings; in particular, hyperons need to be
coupled to scalar mesons weaker than predicted by the SU(6) quark
model.  For certain values of the hyperon--scalar meson couplings
hypernuclear EoS can still produce stellar equilibrium configurations
of compact stars compatible with the two-solar-mass pulsar observations.

Neutrino trapping leads to strong modification in the population of
hyperons and to a shift in the threshold density at which they first appear. As a
consequence, a stiffening of the EoS in the early stage of the neutron
star formation is observed.  Instead of deleptonization with
increasing density, seen in neutrino-less matter, the abundances of
charged leptons remain constant, which among other things leads to
inversion of the density thresholds for appearance of charged
$\Sigma$'s.

\ack
The work of GC was supported by the HGS-HIRe graduate
program at Frankfurt University.

\section*{References}

\providecommand{\newblock}{}


\begin{thebibliography}{10}
\expandafter\ifx\csname url\endcsname\relax
  \def\url#1{{\tt #1}}\fi
\expandafter\ifx\csname urlprefix\endcsname\relax\def\urlprefix{URL }\fi
\providecommand{\eprint}[2][]{\url{#2}}


\bibitem{2010Natur.467.1081D}
{Demorest} P~B, {Pennucci} T, {Ransom} S~M, {Roberts} M~S~E and {Hessels} J~W~T
  2010 {\em \nat\/} {\bf 467} 1081--1083 (\textit{Preprint} \eprint{1010.5788})

\bibitem{2013Sci...340..448A}
{Antoniadis} J, {Freire} P~C~C, {Wex} N, {Tauris} T~M, {Lynch} R~S, {van
  Kerkwijk} M~H, {Kramer} M, {Bassa} C, {Dhillon} V~S, {Driebe} T, {Hessels}
  J~W~T, {Kaspi} V~M, {Kondratiev} V~I, {Langer} N, {Marsh} T~R, {McLaughlin}
  M~A, {Pennucci} T~T, {Ransom} S~M, {Stairs} I~H, {van Leeuwen} J, {Verbiest}
  J~P~W and {Whelan} D~G 2013 {\em Science\/} {\bf 340} 448 (\textit{Preprint}
  \eprint{1304.6875})

\bibitem{CS2013}
{Colucci} G and {Sedrakian} A 2013 {\em \prc\/} {\bf 87} 055806
 (\textit{Preprint} \eprint{1302.6925})


\bibitem{1982PhLB..114..392G}
{Glendenning} N~K 1982 {\em Physics Letters B\/} {\bf 114} 392--396;
{Glendenning} N~K 1985 {\em \apj\/} {\bf 293} 470--493;
{Glendenning} N~K 1987 {\em Zeitschrift fur Physik A Hadrons and Nuclei\/} {\bf
  326} 57--64;
{Glendenning} N~K 1987 {\em Zeitschrift fur Physik A Hadrons and Nuclei\/} {\bf
 f 327} 295--300

\bibitem{1989NuPhA.493..549W}
{Weber} F and {Weigel} M~K 1989 {\em Nuclear Physics A\/} {\bf 493} 549--582

\bibitem{1990PhRvL..64...13K}
{Kapusta} J~I and {Olive} K~A 1990 {\em Physical Review Letters\/} {\bf 64}
  13--15

\bibitem{1991NuPhB.348..345E}
{Ellis} J, {Kapusta} J~I and {Olive} K~A 1991 {\em Nuclear Physics B\/} {\bf
  348} 345--372

\bibitem{1991PhRvL..67.2414G}
{Glendenning} N~K and {Moszkowski} S~A 1991 {\em Physical Review Letters\/}
  {\bf 67} 2414--1417

\bibitem{1995PhLB..349...11E}
{Ellis} P~J, {Knorren} R and {Prakash} M 1995 {\em Physics Letters B\/} {\bf
  349} 11--17 (\textit{Preprint} \eprint{arXiv:nucl-th/9502033})

\bibitem{1996PhRvC..53.1416S}
{Schaffner} J and {Mishustin} I~N 1996 {\em \prc\/} {\bf 53} 1416--1429
  (\textit{Preprint} \eprint{arXiv:nucl-th/9506011})

\bibitem{weber_book}
{Weber} F (ed) 1999 {\em {Pulsars as astrophysical laboratories for nuclear and
  particle physics}\/}

\bibitem{2007PrPNP..58..168S}
{Sedrakian} A 2007 {\em Progress in Particle and Nuclear Physics\/} {\bf 58}
  168--246 (\textit{Preprint} \eprint{arXiv:nucl-th/0601086})

\bibitem{2011PhRvC..84c5801S}
{Schulze} H~J and {Rijken} T 2011 {\em \prc\/} {\bf 84} 035801

\bibitem{2012JPhCS.342a2006L}
{Logoteta} D, {Vida{\~n}a} I, {Provid{\^e}ncia} C, {Polls} A and {Bombaci} I
  2012 {\em Journal of Physics Conference Series\/} {\bf 342} 012006

\bibitem{ring2005}
{Lalazissis} G~A, {Nik{\v s}i{\'c}} T, {Vretenar} D and {Ring} P 2005 {\em
  \prc\/} {\bf 71} 024312

\bibitem{1995PhRvC..52.3043F}
{Fuchs} C, {Lenske} H and {Wolter} H~H 1995 {\em \prc\/} {\bf 52} 3043--3060
  (\textit{Preprint} \eprint{arXiv:nucl-th/9507044})

\bibitem{VMD_model}
{Sakurai} J~J 1960 {\em Annals of Physics\/} {\bf 11} 1--48

\bibitem{2009JHEP...07..105K}
{KLOE Collaboration} 2009 {\em Journal of High Energy Physics\/} {\bf 7} 105
  (\textit{Preprint} \eprint{0906.3819})

\bibitem{de_swart1963}
{de Swart} J~J 1963 {\em Reviews of Modern Physics\/} {\bf 35} 916--939

\bibitem{1989PhRvC..40.2226M}
Maessen P~M~M, Rijken T~A and de~Swart J~J 1989 {\em Phys. Rev. C\/} {\bf
  40}(5) 2226--2245
  \urlprefix\url{http://link.aps.org/doi/10.1103/PhysRevC.40.2226}

\bibitem{2013ApJ...762...96B}
{Bogdanov} S 2013 {\em \apj\/} {\bf 762} 96 (\textit{Preprint}
  \eprint{1211.6113})

\end{thebibliography}
\end{document}